\documentclass[%
reprint,
superscriptaddress,
nobibnotes,
 amsmath,amssymb,
 aps,
prl,
]{revtex4-1}
\usepackage{graphicx}
\usepackage{dcolumn}% Align table columns on decimal point
\usepackage{bm}% bold math
\usepackage[caption=false]{subfig}
\usepackage{floatrow}
\usepackage{ulem}
\usepackage{upgreek}
\usepackage{xcolor}

\definecolor{bluegreen}{HTML}{16b5b2}

\usepackage[colorinlistoftodos]{todonotes}

\begin{document}

\preprint{APS/123-QED}

\title{Phasonic Spectroscopy of a Quantum Gas in a Quasicrystalline Lattice}

\author{Shankari~V.~Rajagopal} 
\author{Toshihiko~Shimasaki}
\author{Peter~Dotti} 
\affiliation{Department of Physics, University of California, Santa Barbara, California 93106, USA}
\author{Mantas~Ra\v{c}i\={u}nas} 
\affiliation{\mbox{Institute of Theoretical Physics and Astronomy, Vilnius University, Saul\.etekio 3, LT-10257 Vilnius, Lithuania}}
\author{Ruwan~Senaratne} 
\affiliation{Department of Physics, University of California, Santa Barbara, California 93106, USA}
\author{Egidijus~Anisimovas}
\affiliation{\mbox{Institute of Theoretical Physics and Astronomy, Vilnius University, Saul\.etekio 3, LT-10257 Vilnius, Lithuania}}
\author{Andr\'e~Eckardt}
\affiliation{Max-Planck-Institut f\"ur Physik komplexer Systeme, N\"othnitzer Str.\ 38, 01187 Dresden, Germany} 
\author{David~M.~Weld}
\email{weld@ucsb.edu}
\affiliation{Department of Physics, University of California, Santa Barbara, California 93106, USA}

\begin{abstract} 
Phasonic degrees of freedom are unique to quasiperiodic structures, and play a central role in poorly-understood properties of quasicrystals from excitation spectra to wavefunction statistics to electronic transport. However, phasons are challenging to access dynamically in the solid state due to their complex long-range  character and the effects of disorder and strain.  
We report phasonic spectroscopy of a  quantum gas in a one-dimensional quasicrystalline optical lattice. We observe that strong phasonic driving produces a nonperturbative high-harmonic plateau strikingly different from the effects of standard dipolar driving. Tuning the potential from crystalline to  quasicrystalline, we identify spectroscopic signatures of quasiperiodicity and interactions and  map the emergence of a multifractal energy spectrum, opening a path to direct imaging of the Hofstadter butterfly. 
\end{abstract}

\maketitle

%\section{BACKGROUND}
Phasons are degrees of freedom unique to quasicrystals~\cite{schechtman-originalQCpaper,phasonreview,bakphasons,steinhardtphononsohasons,Goldman-Widom_AnnualReview}.
The role of phasons in determining quasicrystal properties remains incompletely understood: open questions include the effects of electron-phason coupling, the nature of electronic transport, spectral statistics,  topological properties, and even the shape of the electronic wavefunctions~\cite{hofstadter-fibonacci-butterfly-2007,Thiel-dubois-QCcommentary,QCinAg,Kraus_Zilberberg_Topological_PRB,ZilberbergQC,boundaryphenomena2,verbin-photonic-topological_PRL,brouwerpaper,ames-QCs,hofstadter-superlattice-coldatom-proposal}. These lacunae are in part due to the theoretical intractability of quasiperiodic  matter, and in part due to the experimental difficulty of disentangling the effects of domain walls, crystalline impurities, and disorder from those due to phason modes, which arise from broken translation symmetry in the in the higher-dimensional space from which the quasiperiodic lattice is projected. The exquisite controllability of ultracold atoms in optical lattices makes them well-suited to the study of quasicrystal phenomena from structure to transport to self-similarity~\cite{Lye-Inguscio-interaction-localization_PRA,inguscio-andersonloc,bloch-mbl,Verkerk-Grynberg-5fold_PRL, Verkerk-5fold-diffusion_PRA,QC-corcovilos19,fibonacci-PRA,Gadway-schneble_PRL, Viebahn-Schneider-8fold_PRL,8fold-phase-transition}. Beyond the fundamental interest of such questions, they may point the way to  technological applications of quasicrystals' anomalous electrical and thermal transport characteristics.   
 
Here we report the realization of phasonic spectroscopy on a one-dimensional quasicrystal, using a quantum gas in a tunable bichromatic optical lattice. In addition to standard dipolar modulation, the experiment enables dynamic driving of a phasonic degree of freedom~\cite{Kromer-PhasonTrajectories_PRL,Widom_PhasonReview} via modulation of the relative spatial phase between the two sublattices. We observe that the quasicrystal responds very differently to dipolar and phasonic drives: most strikingly, phasonic modulation generates a broad non-perturbative plateau of high-order ``multi-photon'' transitions, in which multiple energy quanta (``photons'') with energy corresponding to the driving frequency are absorbed. To further elucidate the spectroscopic signatures of quasicrystallinity we measure excitation spectra while varying the strength of quasiperiodicity through a localization transition, observing the emergence of minibands, identifying spectral features arising from interatomic interactions and localization-induced diabaticity, and  mapping a slice of the Hofstadter butterfly energy spectrum.  

\begin{figure}[t]
\centering
\includegraphics[width=\columnwidth]{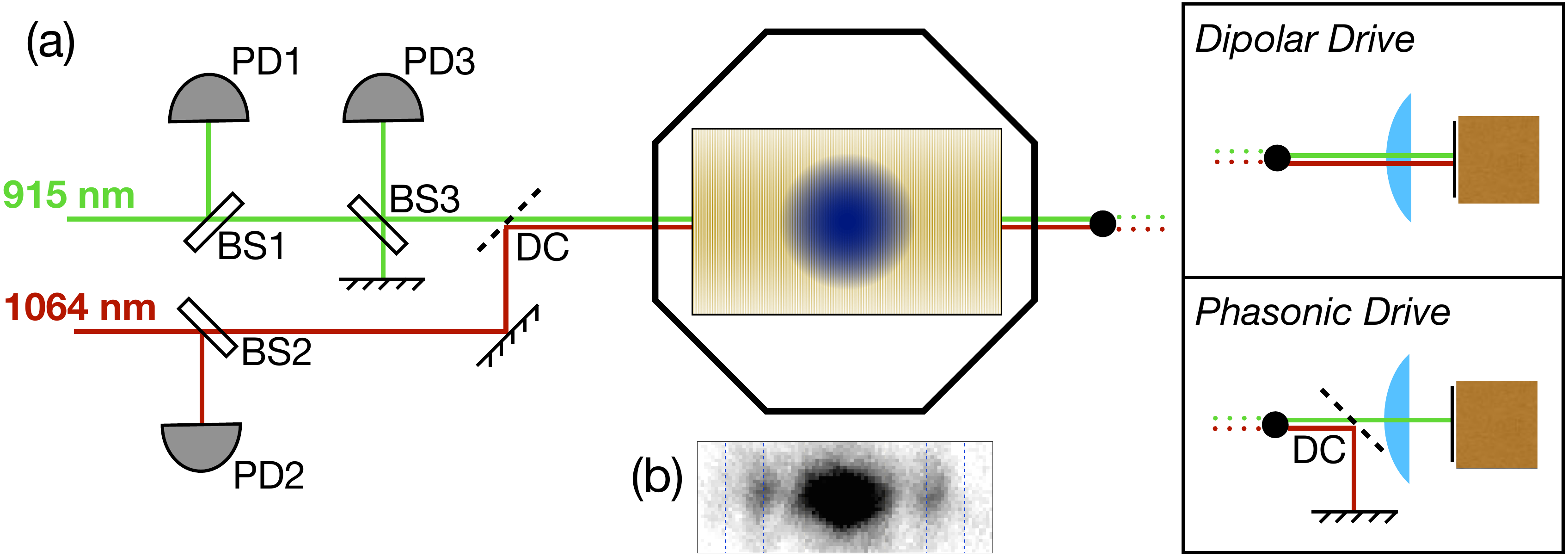}
\caption{Experimental schematic. (a)BEC (blue) in a bichromatic lattice (yellow). Photodiodes (PD), beam samplers (BS), and dichroic mirrors (DC) are indicated, as is the configuration for both dipolar and phasonic driving using a piezo-driven mirror (solid block).   (b) Sample band-mapped data. Dotted lines indicate zone edges of the primary lattice.}
\label{fig:fig0}
\end{figure}

%\section{EXPERIMENT}

\begin{figure*}[t!]
\includegraphics[width=\textwidth]{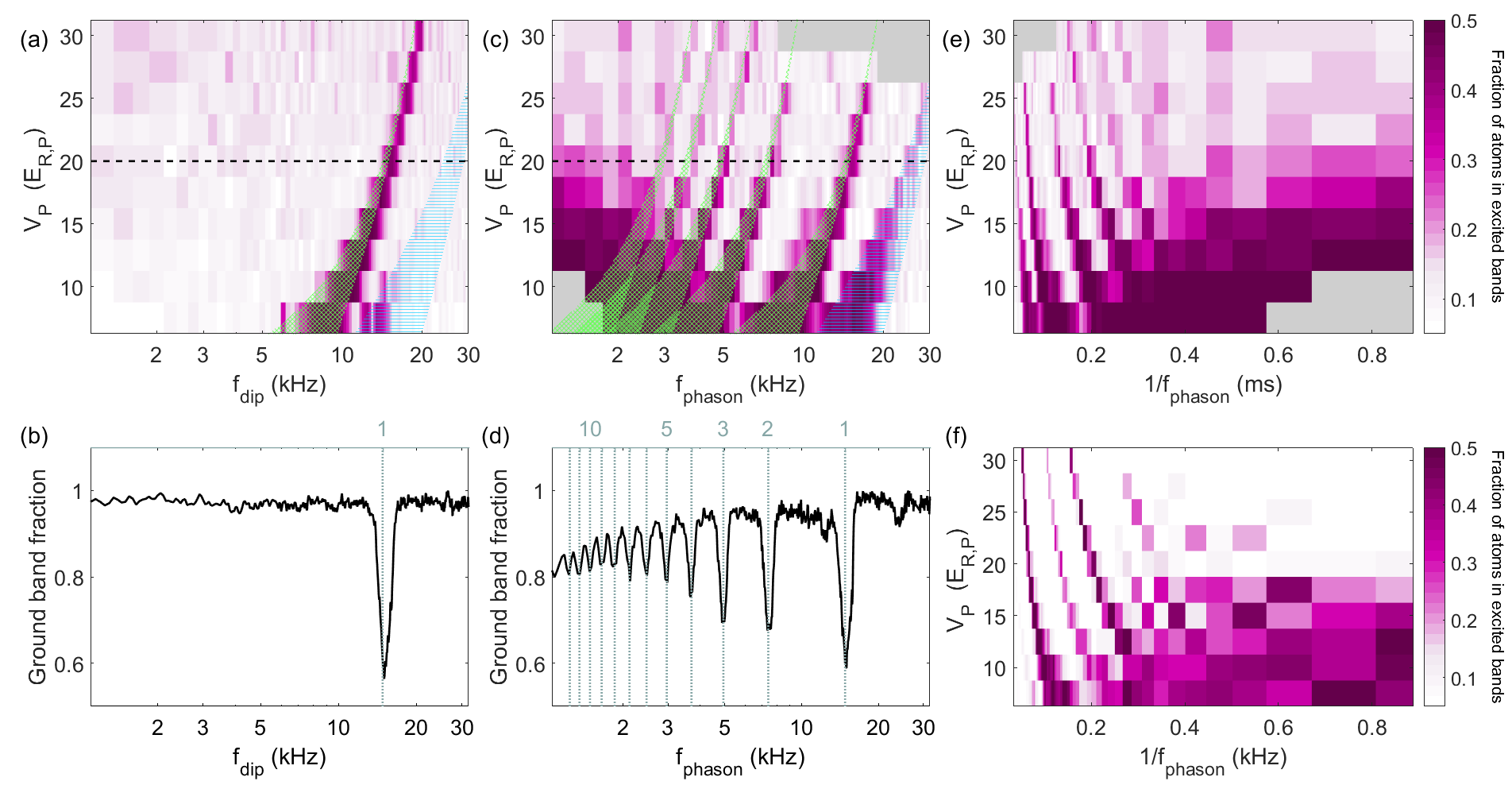}
\caption{Comparison of dipolar and phasonic spectroscopy; areas where no data was taken are marked in gray. \textbf{(a)} Excitation due to dipolar driving as a function of drive frequency $f_\mathrm{dip}$ and primary lattice depth $V_P$ with $\alpha_{\mathrm{dip}}=0.16\times V_S/V_P$ and
% formerly $\alpha_{\mathrm{dip}}=0.22\times E_{R,P}/V_P$
 $V_S=1 E_{R,S}$. Green hatched (Blue horizontal) overlay shows calculated first (second) interband transition. \textbf{(b)} High-resolution dipolar spectrum at $V_P=20 E_{R,P}$. Line shows the calculated center of the first interband transition.  \textbf{(c)} Excitation due to phasonic driving as a function of drive frequency $f_\mathrm{phason}$ and primary lattice depth $V_P$. $\alpha_{\mathrm{phason}}$ is set to $\approx$ 1. Green hatched (Blue horizontal) overlays show calculated first (second) interband transition, with multiphoton subharmonics also indicated for the first transition. \textbf{(d)} High-resolution phasonic spectrum at $V_P=20 E_{R,P}$. Lines show the calculated center of the first twelve multiphoton transitions corresponding to the lowest interband transition. \textbf{(e)} Data from (c) plotted versus drive period $1/f_\mathrm{phason}$, showing a broad low-frequency absorption feature. \textbf{(f)} Theoretical prediction for (e) (details in text).}
\label{fig:fig1}
\end{figure*}

The experiments (diagrammed in Fig.~\ref{fig:fig0}) use a 1D bichromatic potential which superposes a primary  and secondary lattice formed by light with wavelengths $\lambda_P = $ 1064 nm and $\lambda_S = $ 915 nm. Neglecting interactions, the  Hamiltonian of atoms in this potential is 

\begin{equation}
\begin{aligned}
H ={}  -\frac{\hbar^2}{2m} \frac{d^2}{dx^2} & + \frac{V_P}{2} \cos(2k_P(x-\delta_P)) \\[1.2ex]
                & +\frac{V_S}{2} \cos(2k_S(x-\delta_S)),
\end{aligned}
\label{eq:aamodel}
\end{equation}
% \begin{equation}
% \begin{aligned}
% H =-\frac{\hbar^2}{2m}\frac{d^2}{dx^2}\!+\!\frac{V_P}{2}\!\cos(2k_P(x\!-\!\delta_P))\!+\!\frac{V_S}{2}\!\cos(2k_S(x\!-\!\delta_S)),
% \end{aligned}
% \label{eq:aamodel}
% \end{equation}
where $k_{P(S)} = 2\pi/\lambda_{P(S)}$ and $V_P$ and $\delta_P$ ($V_S$ and $\delta_S$) are the amplitude and spatial phase of the primary (secondary) lattice. For $V_P\gg V_S$, in the tight-binding limit with respect to the primary lattice, this Hamiltonian is closely related to both the Aubry-Andr\'e model~\cite{aubryandre} and the Harper model~\cite{harpermodel}; for larger $V_S$, deviations from these models appear in the form of mobility edges~\cite{spme-boers_PRA07,spme-das-sarma_PRB,spme-bloch_PRL18,monika-mbl-spme}. Lattice depths are measured in the respective recoil energies, $E_{R,i} = \hbar^2 k_i^2/2m$, $i\!\in\!\{P,S\}$. The chosen value of the ratio $\nu = \lambda_S/\lambda_P$ is effectively irrational in the sense that it gives rise to a unit cell larger than our 30 $\upmu$m sample size; in other words, the potential is quasiperiodic to within experimental resolution.

A key feature of the experiment is the ability to modulate the different degrees of freedom of the bichromatic lattice. Standard dipolar excitation, which drives the lowest-energy phononic mode of the lattice, is achieved by equal translation of both lattices:
\begin{equation}
\delta_S(t)\ =\ \delta_P(t)\ =\ A_\mathrm{dip} \sin(2\pi f_\mathrm{dip} t),
\label{phononeq}
\end{equation}
% \begin{equation}
% \begin{aligned}
% \delta_S(t) &= A_\mathrm{phonon} \sin(2\pi f_\mathrm{phonon} t),\\
% \delta_P(t) &= A_\mathrm{phonon} \sin(2\pi f_\mathrm{phonon} t).
% \end{aligned}
% \label{phononeq}
% \end{equation}
\noindent where $A_\mathrm{dip}$ and $f_\mathrm{dip}=\omega_\mathrm{dip}/2\pi$ are the amplitude and frequency of the dipolar drive. In the lattice frame the force applied to the atoms is $F(t) = F_0 \sin({2\pi f_\mathrm{dip} t)}$ for $F_0 = m (2\pi f_\mathrm{dip})^2 A_\mathrm{dip}$. Using the primary lattice constant $a=\lambda_P/2$, we define a dimensionless driving parameter % \begin{equation}
% \begin{aligned}
% K&= a F_0/\hbar \omega_\mathrm{phonon}\\
% &= a m \omega_\mathrm{phonon} A_\mathrm{phonon}/\hbar;
% \end{aligned}
% \label{eq:dimlessamp}
% \end{equation}
$\alpha_\mathrm{dip} = a F_0/\hbar \omega_\mathrm{dip} = a m \omega_\mathrm{dip} A_\mathrm{dip}/\hbar$ (which determines the modification of tunnelling matrix elements in the lowest band~\cite{Eckardt2017}).
To keep $\alpha_\mathrm{dip}$ fixed for different drive frequencies, we take $A_\mathrm{dip} \propto 1/f_\mathrm{dip}$; this normalization procedure for phase modulation has been used previously to study multiphoton excitations in a single-color lattice~\cite{interband_eckardt_sengstock_simonet, latticeheating-blochgroup}. Phasonic modulation is achieved by translating only the secondary lattice: 
\begin{equation}
%\begin{aligned}
\delta_S(t) = A_\mathrm{phason} \sin(2\pi f_\mathrm{phason} t),\,\,\,\,\,\,\,\,
\delta_P(t) = 0,
%\end{aligned}
\label{phasoneq}
\end{equation}
\noindent where $A_\mathrm{phason}$ and $f_\mathrm{phason}$ are the  amplitude and frequency of the phasonic drive. As with the dipolar drive, we define a dimensionless amplitude $\alpha_\mathrm{phason}$, for which $A_\mathrm{phason} =C\alpha_\text{phason}/f_\text{phason}$, taking $C=1000 \text{ nm}\cdot\text{kHz}$. %Note that, unlike the inertial dipolar force, the phasonic drive produces matrix elements $\propto \alpha_{\mathrm{phason}}V_S$ that are proportional to the depth of the secondary lattice. 

%To enable the comparison of dipolar and phasonic driving shown in Fig.~2, either the overall force or the observed rate of resonant transitions to the first excited band can be held constant; these procedures give similar values of $A_\mathrm{phason}/A_\mathrm{dip}\simeq V_S/V_P$. SHANKARI PLEASE CHECK THIS SENTENCE FOR ACCURACY-- IT IS A KEY POINT.

%to keep a dimensionless driving amplitude fixed [PD: How sure are we that a so defined dimensionless driving is appropriate?  Andre seemed skeptical.][SVR: Not sure at all, but it is a description of what we do for the datasets].  [PD:  I suppose its not untrue to use this language, but I believe describing it as a dimensionless drive parameter somehow implies that it simplifies the theory to use this parameter.  Perhaps this can be shown, but if we don't, I think I would prefer saying something like "we scale the drive for a more direct comparison to the phononic drive". Or something.]

We report spectroscopic measurements of the quasicrystal's response to phasonic and dipolar excitation with varying drive and lattice parameters. The experiments begin by adiabatically loading a Bose condensate of $^{84}$Sr into the bichromatic lattice. The amplitude of dipolar or phasonic modulation is linearly ramped to the final value over 4~ms, followed by constant-amplitude modulation for 16~ms. After modulation, both lattices are ramped down simultaneously at a rate which is adiabatic with respect to the energy gaps of the primary lattice, to achieve approximate band mapping onto free-space momentum states~\cite{demarco-bandmap}. This enables measurement of the primary spectroscopic observable: the fractional population of atoms in the ground band of the primary lattice after modulation.  

%\begin{enumerate}
%\item Intro to system [IS A FIGURE OF EXPTL SETUP NECESSARY HERE?]. Hamiltonian, etc.
%\item Discussion of hamiltonian and general driving parameters for phonons/phasons 
%\item Discussion of our system and our driving parameters in terms of the general ones in the hamiltonian (we need to come up with some generalized drive parameter which normalizes to pulse length, etc)
%\end{enumerate}

%\section{Phasonic Spectroscopy}

%%%% SVR: Fig 1 has been placed at the beginning of Experiment section for proper placement in paper. It can be moved back if appropriate.

%Main points and punchlines: The phasonic degree of freedom of quasicrystals is interesting and unexplored. We perform phasonic spectroscopy of a quantum gas in a quasiperiodic lattice.  We observe that phasonic driving excites high-order multiphoton modes more efficiently than phononic driving.  We observe a critical driving amplitude for these multiphoton transitions which depends weakly on transition order (more detail needed from theorists).  We observe that phasonic driving at large tunneling strengths and low frequencies excites a broad continuum of modes, in contrast to phononic driving under the same conditions (not sure we can say this-- depends on results of blob amplitude dependence data set.  But we should say something about the blob.).

As a first application of phasonic spectroscopy, we measure and plot in Fig.~2 the difference between a quasicrystal's response to standard dipolar driving and its response to phasonic driving. %To facilitate comparison we use comparable fixed dimensionless drive amplitudes:  $\alpha_{\mathrm{dip}}=0.22\times E_{R,P}/V_P$, with the phasonic drive amplitude scaled up by $V_P/V_S$ such that the ground band depletion fraction of the fundamental interband resonance is approximately  the same for both drives. For these measurements we vary the primary lattice depth and modulation frequency, holding the secondary lattice depth fixed at 1~$E_{R,S}$. 
We fix the phasonic driving amplitude to $\alpha_{\mathrm{phason}}\approx 1$. To facilitate comparison, the dipolar drive is scaled with respect to the phasonic one by a factor proportional to the sublattice depth ratio: $\alpha_\mathrm{dip} = 0.16\times V_S/V_P$. $V_S$ is held at $1E_{R,S}$ for both drives.
%we scale driving amplitudes accordingly. For the phasonic driving, we fix the driving amplitude to $\alpha_\mathrm{phason}\approx 1$. For the dipolar drive, the amplitude is set as $\alpha_\mathrm{dip} = 0.22\times E_{R,P}/V_P$, giving a scaling factor of $\sim V_S/V_P$. % with respect to the phasonic one, $\alpha_\text{dip}\approx 0.22\times E_{R,P}/V_P$, while $V_S$ is held fixed at $1E_{R,S}$ for both drives.
Dipolar driving causes excitations to higher bands which are  consistent with expected interband transitions of the primary lattice (Fig.~\ref{fig:fig1}(a) and \ref{fig:fig1}(b)). The second interband transition is visible but suppressed compared to the first transition, since the odd-parity dipolar force does not couple unperturbed Wannier states of the primary lattice of equal parity on the same site.  No multiphoton transitions are apparent at this drive amplitude. 

The response to phasonic driving is qualitatively different, although due to the chosen $\alpha$ scaling the main interband transition is driven at similar strength. Most strikingly, we observe strong multiphoton processes up to the twelfth order (Fig.~\ref{fig:fig1}(c) and \ref{fig:fig1}(d)). Phasonic excitation in this regime apparently gives rise to an efficient high-harmonic response, in which atoms can absorb energy at high multiples of the drive frequency.  Additionally, comparison of Fig.~2(a) and 2(c) indicates that phasonic driving appears to relax the suppression of even interband transitions, which we attribute to the fact that it is not parity (anti)symmetric on site. Finally, we observe a broad low-frequency absorption feature at large tunneling amplitudes in the phasonic spectrum. This feature, most easily seen in Fig.~\ref{fig:fig1}(e), is likely due to overlap of numerous high-order harmonics. The experimental results in Fig.~\ref{fig:fig1}(e) are reproduced well by the non-interacting exact time-evolution numerical simulations shown in Fig.~\ref{fig:fig1}(f)~\cite{SuppMat}.  The experimental observation of these unique features of phasonic spectroscopy of a tunable quantum quasicrystal --- efficient high-harmonic response, relaxed selection rules, and broadband IR absorption --- constitute the first main result of this report. 

\begin{figure}[t]
\centering
\includegraphics[width=\columnwidth]{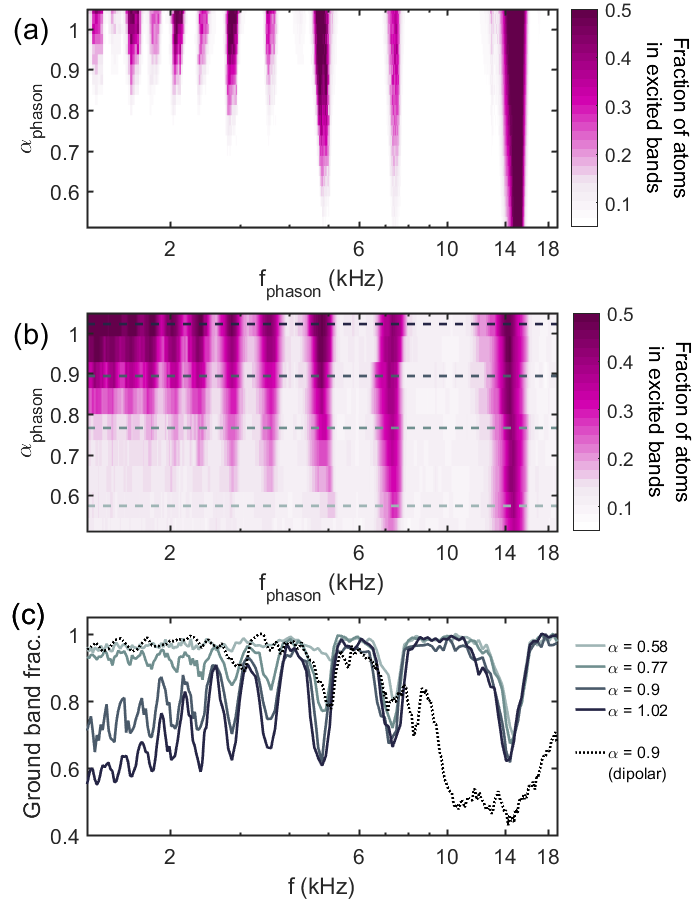}
\caption{Amplitude dependence of multiphoton resonances. (a) Theoretical simulation of phasonic spectra for varying drive amplitude $\alpha_\mathrm{phason}$. (b) Experimentally measured phasonic spectra for $V_P=20 E_{R,P}$ and varying $\alpha_\mathrm{phason}$. Both experiment and theory show the onset of a non-perturbative regime near $\alpha_\mathrm{th} = 0.9$. (c) Line cuts of experimental phasonic (solid) and dipolar (dashed) spectra at various $\alpha$ values. Note the extreme power broadening in the dipolar spectrum.}
\label{fig:fig2}
\end{figure}

\begin{figure*}[t!]
\centering
\includegraphics[width=\columnwidth]{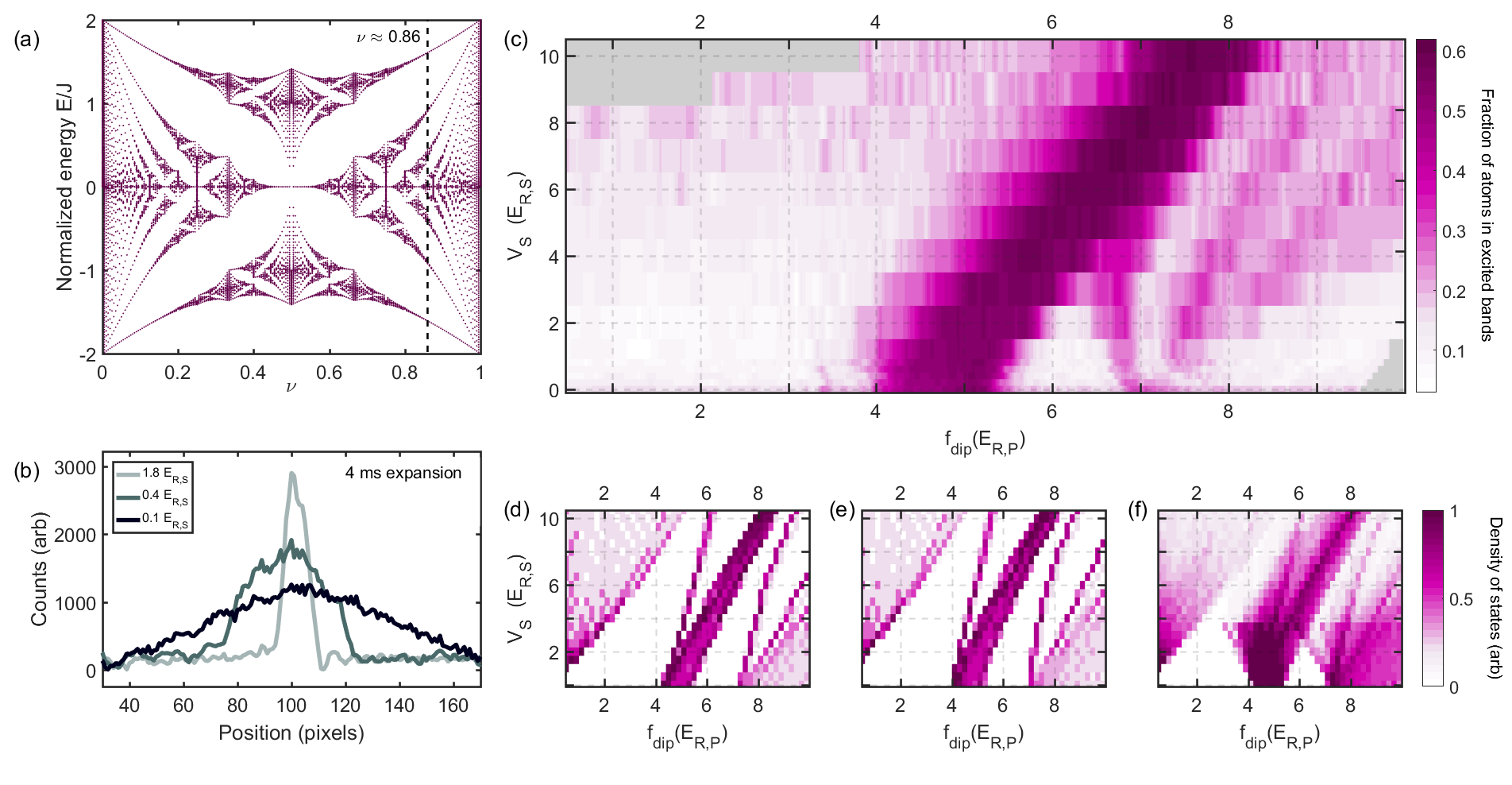}
%\sidesubfloat[]{
%    \includegraphics[width=0.95\columnwidth]{phason915depth.png}
%    \label{fig:fig3b}
%}
\caption{Spectroscopy of an interacting quasicrystal. (a) Calculated energy spectrum vs. $\nu=\lambda_S/\lambda_P$. Dashed line shows the slice corresponding to the quasicrystal used in this experiment. (b) Post-expansion atomic density distribution at varying disorder strengths $V_S$, showing the effects of crossing the localization transition. (c) Experimentally measured dipolar excitation spectra for varying $V_S/V_P$ at $\alpha_{\mathrm{dip}}=0.022$, showing spectral minigaps. No data were taken for the gray areas in the upper-left and lower-right. (d) Calculated density of final states for a non-interacting system, starting from a BEC. (e) Calculated density of final states for an interacting BEC; a shift of the resonance line to lower frequencies from Fig.~\ref{fig:fig3}(d) is observed. (f) Calculated non-interacting transition density assuming all single-particle orbitals below 1.5 $E_{R,S}$ are initially populated.}
\label{fig:fig3}
\end{figure*}

The proliferation of multiphoton resonances is connected to the breakdown of the regime where the driving amplitude can be treated perturbatively \cite{Straeter2016}. In the phasonically excited quasicrystal, the threshold for entering the non-perturbative regime can be estimated by expanding the shaken secondary lattice potential as
$\cos(2k_S [x-A\sin(\omega t)])=\sum_{n=-\infty}^\infty J_n(2k_SA)[\cos(2 k_S x)\cos(n\omega t)+\sin(2 k_S x)\sin(n\omega t)]$, where the Bessel functions $J_n$ contain all powers (orders) of the scaled driving amplitude 
%$2k_S A=(4\nu E_{R,S}/\pi\hbar\omega)\alpha_\mathrm{phason}$. 
$2k_S A = (4 \pi k_S C / \omega)\alpha_\mathrm{phason}$. 
The individual terms can directly induce $n$-photon transitions with $\Delta E = |n|\hbar\omega$, but (as a property of the Bessel function) contribute only as long as $|n|\lesssim 2k_SA$, corresponding to the estimated threshold value 
%$\alpha_\text{th}=\pi\Delta E/ 4\nu E_{R,S}$ 
$\alpha_\text{th} = \Delta E /  4\pi \hbar k_S C$ 
for the dimensionless driving amplitude $\alpha_\text{phason}$. For transitions to the first excited band, we obtain 
%$\alpha_\text{th}\approx 0.9$, 
$\alpha_\text{th}$ close to unity~\cite{SuppMat},  
in reasonable agreement with both the numerical simulations shown in Fig~3(a) and the experimental data shown in Fig.~3(b). 
%For phasonic spectroscopy at fixed quasicrystal parameters but varying dimensionless drive amplitude $\alpha$, the numerical calculations shown in Fig.~3a and the analytic approximations discussed in the Supplemental Material both predict a critical phasonic drive amplitude in the neighborhood of $\alpha=0.9$, above which a plateau of multiphoton resonances emerges.
%Experimental spectra taken for varying $\alpha$, plotted in Fig.~3b and 3c, are in good agreement these predictions. 
Note in particular that at low frequencies, in order to keep $\alpha_\mathrm{phason}$ constant and equivalent to that used for the dipolar drive, the position-space amplitude of the phasonic drive used for the data shown in Fig.~2 increases to significantly more than one lattice constant. A noticeable difference between the dipolar and the phasonic drive is the significantly flatter distribution of transition strengths in the phasonic case.  As an experimental indication of this effect, note the comparison of dipolar and phasonic spectra in Fig.~3(c): while the high-amplitude phasonic spectrum shows numerous narrow transition lines, a dipolar spectrum taken at an amplitude sufficient to weakly drive multiphoton transitions already exhibits extreme power-broadening of the first interband transition, in agreement with previous work on periodic lattices~ \cite{interband_eckardt_sengstock_simonet,latticeheating-blochgroup}. As an additional point of interest, we note that connections between harmonic generation and quasiperiodicity have been made in other physical systems~\cite{physletta-phonon-limas05,science-hhgqc-zhu97}.

The spectroscopic probes of tunable cold-atom quasicrystals which we demonstrate can be deployed to study the rich variety of phenomena arising from quasiperiodicity and interactions. As a consequence of a  mapping to the Harper model~\cite{harpermodel}, the non-interacting energy spectrum of a 1D quasicrystal constitutes a slice through the multifractal spectrum of two-dimensional electron gases in the integer quantum Hall regime known as the Hofstadter butterfly~\cite{hofstadter-theoriginal}, plotted in Fig.~4(a). Mapping this fascinating spectrum in an entirely different physical context than high-field 2D Fermi gases, and probing the interplay of interactions and quasiperiodicity are two natural applications for the spectroscopic techniques we describe.

%\section{Tuning quasiperiodicity}

With these goals in mind, we  measured the evolution of the spectral response as the strength of quasiperiodicity was increased from zero by tuning $V_S/V_P$. While the flattened selection rules of phasonic driving are potentially appealing for such a measurement, phasonic driving is not available at $V_S/V_P=0$ and in any case the strong high-harmonic response complicates the interpretation of phasonic spectra. Therefore for this measurement we chose to use  dipolar driving.  Tuning of quasiperiodicity was achieved for a fixed $V_P$ by varying the relative strength of the weaker lattice $V_S/V_P$ between zero and one. We note that this range spans a localization transition of the generalized Aubry-Andr\'e type~\cite{aubryandre,inguscio-andersonloc} which has strong effects on transport: Fig.~4(b) shows atomic density distributions after 4 ms of expansion at various values of $V_S/V_P$, clearly indicating the effects of localization. Fig.~4(c) shows results of dipolar modulation spectroscopy on a 10~$E_{R,P}$ primary lattice at fixed $\alpha_\mathrm{dip} = 0.022$ and variable $V_S$, allowing direct measurement of the spectral effects of bichromaticity. We observe the formation of the ``minigaps'' which are hallmarks of the Hofstadter spectrum. 

Comparison of the experimental data (Fig.~\ref{fig:fig3}(c)) to the theoretically computed density of states for the non-interacting quasiperiodic lattice (Fig.~\ref{fig:fig3}(d)) reveals a number of interesting features. For small $V_S$, we can clearly identify the three lowermost bands in Fig.~\ref{fig:fig3}(d). While the ground band is not reflected in the experimental data, since intraband excitations were not measured, the observed main resonance clearly corresponds to excitations to the first excited band. However, the blueshift (bending to the right) of this resonance with increasing $V_S$ is clearly lower in the experimental data than in the theory of Fig.~\ref{fig:fig3}(d). We attribute this effect to interactions; taking them into account on a mean-field level~\cite{SuppMat}, we obtain a reduced blueshift in agreement with experiment (Fig.~\ref{fig:fig3}(e)). Excitations to the second excited band are suppressed by weak coupling matrix elements, since for $V_S=0$ the dipolar drive couples on-site Wannier states of opposite parity. Nevertheless, the experiment shows a few narrow resonance lines, induced by switching on a finite $V_S$; we interpret these as a signature of the emergence of minibands. The experimental plot also features an additional resonance line, which merges with the main resonance near $V_S/E_{R,S}=4$. This feature can be reproduced by a theory which includes the initial presence of excitations as a result of localization-induced non-adiabatic loading (Fig.~\ref{fig:fig3}(f))~\cite{SuppMat}. This line is effectively a copy of the resonance immediately to its right, corresponding to transitions into these states from the now populated upper edge of the ground band where the density of states shows a pronounced peak (Fig.~\ref{fig:fig3}(d)).  The observation of the emergence of minigaps in a slice of the Hofstadter butterfly spectrum and the identification of spectral shifts due to interactions in a quasicrystal together constitute the second main result of this report.

%\section{CONCLUSION AND OUTLOOK}
The techniques and results we present open up several exciting directions for future work. Most broadly, they  enable  exploration of numerous open questions concerning quantum quasicrystals. Continuous tuning of the period ratio of the bichromatic lattice would allow direct mapping of the 2D Hofstadter butterfly spectrum. Spectroscopy across the Aubry-Andr\'e transition may allow study of the effects of localization on heating processes, including in regions with band-dependent localization and single-particle mobility edges~\cite{spme-boers_PRA07,spme-das-sarma_PRB,spme-bloch_PRL18,monika-mbl-spme}. Monotonic increase of a phasonic degree of freedom  should allow the realization of various topological pumps: a recent proposal suggests high-temperature topological quantized phasonic Thouless pumping of bulk states~\cite{Lindner-Rudner-thouless_PRX}, and the Hofstadter spectrum supports edge states which can be topologically pumped from one end of the system to the other in a single phasonic cycle~\cite{topological-quasicrystals-PRL,fibonacci-PRA}.

In conclusion, we have demonstrated phasonic spectroscopy of a tunable quantum quasicrystal, showed theoretically and experimentally  that phasonic excitation  efficiently drives non-perturbative high-order multiphoton processes and gives rise to a broad low-frequency absorption feature, mapped the spectral features of a transition from a crystal with extended states to a quasicrystal with localized states, measured the emergence of minigaps in a slice of the Hofstadter spectrum, and identified spectral shifts due to the presence of interactions in a quasicrystal.

\begin{acknowledgements}
The authors thank Zach Geiger, Cora Fujiwara, Kevin Singh, and Max Prichard for experimental assistance and G.~\v{Z}labys and U. Schneider for useful discussions. DW acknowledges support from the National Science Foundation (CAREER 1555313), the Office of Naval Research (N00014-16-1-2225), the Army Research Office (PECASE W911NF1410154 and MURI W911NF1710323), and the University of California's Multicampus Research Programs and Initiatives (MRP-19-601445). The work of MR and EA was supported by the European Social Fund under Grant No. 09.3.3-LMT-K-712-01-0051. AE acknowledges funding by the Deutsche Forschungsgemeinschaft (DFG) via the Research Unit FOR 2414 under Project No.\ 277974659.
\end{acknowledgements}

%\begin{enumerate}
%\item We have demonstrated phasonic spectroscopy of a tunable quantum quasicrystal. We can 
%\item In the future this setup presents interesting possibilities to study topological pumping of predicted edge states. Also opens up directions to study Anderson localization in many dimensions. Etc.
%\end{enumerate}

% \section{TO-DO}
% (this is a pretty old list, needs gardening)
% \begin{itemize}
% \item satisfy ourselves re: lattice alignment and aa transition point
% \item retake 1d spec phasonically and phononically with 1d trap nodts
% \item check a couple of amplitudon things
% \item add hats to path and clean up imaging for better spectrum resolution
% \item try a different wavelength
% \item do some numerics
% \item slot some existing figures/comparisons into this overleaf
% \item start fleshing out some sections which we know will eventually exist
% \item solidify theoretical understanding of experiment 
% \end{itemize}
%\bibliography{dmwbibliography,svr_bib}

\begin{thebibliography}{44}%
\makeatletter
\providecommand \@ifxundefined [1]{%
 \@ifx{#1\undefined}
}%
\providecommand \@ifnum [1]{%
 \ifnum #1\expandafter \@firstoftwo
 \else \expandafter \@secondoftwo
 \fi
}%
\providecommand \@ifx [1]{%
 \ifx #1\expandafter \@firstoftwo
 \else \expandafter \@secondoftwo
 \fi
}%
\providecommand \natexlab [1]{#1}%
\providecommand \enquote  [1]{``#1''}%
\providecommand \bibnamefont  [1]{#1}%
\providecommand \bibfnamefont [1]{#1}%
\providecommand \citenamefont [1]{#1}%
\providecommand \href@noop [0]{\@secondoftwo}%
\providecommand \href [0]{\begingroup \@sanitize@url \@href}%
\providecommand \@href[1]{\@@startlink{#1}\@@href}%
\providecommand \@@href[1]{\endgroup#1\@@endlink}%
\providecommand \@sanitize@url [0]{\catcode `\\12\catcode `\$12\catcode
  `\&12\catcode `\#12\catcode `\^12\catcode `\_12\catcode `\%12\relax}%
\providecommand \@@startlink[1]{}%
\providecommand \@@endlink[0]{}%
\providecommand \url  [0]{\begingroup\@sanitize@url \@url }%
\providecommand \@url [1]{\endgroup\@href {#1}{\urlprefix }}%
\providecommand \urlprefix  [0]{URL }%
\providecommand \Eprint [0]{\href }%
\providecommand \doibase [0]{http://dx.doi.org/}%
\providecommand \selectlanguage [0]{\@gobble}%
\providecommand \bibinfo  [0]{\@secondoftwo}%
\providecommand \bibfield  [0]{\@secondoftwo}%
\providecommand \translation [1]{[#1]}%
\providecommand \BibitemOpen [0]{}%
\providecommand \bibitemStop [0]{}%
\providecommand \bibitemNoStop [0]{.\EOS\space}%
\providecommand \EOS [0]{\spacefactor3000\relax}%
\providecommand \BibitemShut  [1]{\csname bibitem#1\endcsname}%
\let\auto@bib@innerbib\@empty
%</preamble>
\bibitem [{\citenamefont {Shechtman}\ \emph {et~al.}(1984)\citenamefont
  {Shechtman}, \citenamefont {Blech}, \citenamefont {Gratias},\ and\
  \citenamefont {Cahn}}]{schechtman-originalQCpaper}%
  \BibitemOpen
  \bibfield  {author} {\bibinfo {author} {\bibfnamefont {D.}~\bibnamefont
  {Shechtman}}, \bibinfo {author} {\bibfnamefont {I.}~\bibnamefont {Blech}},
  \bibinfo {author} {\bibfnamefont {D.}~\bibnamefont {Gratias}}, \ and\
  \bibinfo {author} {\bibfnamefont {J.~W.}\ \bibnamefont {Cahn}},\ }\href
  {\doibase 10.1103/PhysRevLett.53.1951} {\bibfield  {journal} {\bibinfo
  {journal} {Phys. Rev. Lett.}\ }\textbf {\bibinfo {volume} {53}},\ \bibinfo
  {pages} {1951} (\bibinfo {year} {1984})}\BibitemShut {NoStop}%
\bibitem [{\citenamefont {de~Boissieu}(2012)}]{phasonreview}%
  \BibitemOpen
  \bibfield  {author} {\bibinfo {author} {\bibfnamefont {M.}~\bibnamefont
  {de~Boissieu}},\ }\href@noop {} {\bibfield  {journal} {\bibinfo  {journal}
  {Chem. Soc. Rev.}\ }\textbf {\bibinfo {volume} {41}},\ \bibinfo {pages}
  {6778} (\bibinfo {year} {2012})}\BibitemShut {NoStop}%
\bibitem [{\citenamefont {Bak}(1985)}]{bakphasons}%
  \BibitemOpen
  \bibfield  {author} {\bibinfo {author} {\bibfnamefont {P.}~\bibnamefont
  {Bak}},\ }\href {\doibase 10.1103/PhysRevB.32.5764} {\bibfield  {journal}
  {\bibinfo  {journal} {Phys. Rev. B}\ }\textbf {\bibinfo {volume} {32}},\
  \bibinfo {pages} {5764} (\bibinfo {year} {1985})}\BibitemShut {NoStop}%
\bibitem [{\citenamefont {Socolar}\ \emph {et~al.}(1986)\citenamefont
  {Socolar}, \citenamefont {Lubensky},\ and\ \citenamefont
  {Steinhardt}}]{steinhardtphononsohasons}%
  \BibitemOpen
  \bibfield  {author} {\bibinfo {author} {\bibfnamefont {J.~E.~S.}\
  \bibnamefont {Socolar}}, \bibinfo {author} {\bibfnamefont {T.~C.}\
  \bibnamefont {Lubensky}}, \ and\ \bibinfo {author} {\bibfnamefont {P.~J.}\
  \bibnamefont {Steinhardt}},\ }\href {\doibase 10.1103/PhysRevB.34.3345}
  {\bibfield  {journal} {\bibinfo  {journal} {Phys. Rev. B}\ }\textbf {\bibinfo
  {volume} {34}},\ \bibinfo {pages} {3345} (\bibinfo {year}
  {1986})}\BibitemShut {NoStop}%
\bibitem [{\citenamefont {Goldman}\ and\ \citenamefont
  {Widom}(1991)}]{Goldman-Widom_AnnualReview}%
  \BibitemOpen
  \bibfield  {author} {\bibinfo {author} {\bibfnamefont {A.~I.}\ \bibnamefont
  {Goldman}}\ and\ \bibinfo {author} {\bibfnamefont {M.}~\bibnamefont
  {Widom}},\ }\href@noop {} {\bibfield  {journal} {\bibinfo  {journal} {Annual
  Review of Physical Chemistry}\ }\textbf {\bibinfo {volume} {42}},\ \bibinfo
  {pages} {685} (\bibinfo {year} {1991})}\BibitemShut {NoStop}%
\bibitem [{\citenamefont {Naumis}\ and\ \citenamefont
  {Lopez-Rodriguez}(2008)}]{hofstadter-fibonacci-butterfly-2007}%
  \BibitemOpen
  \bibfield  {author} {\bibinfo {author} {\bibfnamefont {G.~G.}\ \bibnamefont
  {Naumis}}\ and\ \bibinfo {author} {\bibfnamefont {F.}~\bibnamefont
  {Lopez-Rodriguez}},\ }\href {\doibase 10.1016/j.physb.2007.10.016} {\bibfield
   {journal} {\bibinfo  {journal} {Physica B: Condensed Matter}\ }\textbf
  {\bibinfo {volume} {403}},\ \bibinfo {pages} {1755 } (\bibinfo {year}
  {2008})}\BibitemShut {NoStop}%
\bibitem [{\citenamefont {Thiel}\ and\ \citenamefont
  {Dubois}(2000)}]{Thiel-dubois-QCcommentary}%
  \BibitemOpen
  \bibfield  {author} {\bibinfo {author} {\bibfnamefont {P.~A.}\ \bibnamefont
  {Thiel}}\ and\ \bibinfo {author} {\bibfnamefont {J.~M.}\ \bibnamefont
  {Dubois}},\ }\href {http://dx.doi.org/10.1038/35020657} {\bibfield  {journal}
  {\bibinfo  {journal} {Nature}\ }\textbf {\bibinfo {volume} {406}},\ \bibinfo
  {pages} {570} (\bibinfo {year} {2000})}\BibitemShut {NoStop}%
\bibitem [{\citenamefont {Moras}\ \emph {et~al.}(2006)\citenamefont {Moras},
  \citenamefont {Theis}, \citenamefont {Ferrari}, \citenamefont {Gardonio},
  \citenamefont {Fujii}, \citenamefont {Horn},\ and\ \citenamefont
  {Carbone}}]{QCinAg}%
  \BibitemOpen
  \bibfield  {author} {\bibinfo {author} {\bibfnamefont {P.}~\bibnamefont
  {Moras}}, \bibinfo {author} {\bibfnamefont {W.}~\bibnamefont {Theis}},
  \bibinfo {author} {\bibfnamefont {L.}~\bibnamefont {Ferrari}}, \bibinfo
  {author} {\bibfnamefont {S.}~\bibnamefont {Gardonio}}, \bibinfo {author}
  {\bibfnamefont {J.}~\bibnamefont {Fujii}}, \bibinfo {author} {\bibfnamefont
  {K.}~\bibnamefont {Horn}}, \ and\ \bibinfo {author} {\bibfnamefont
  {C.}~\bibnamefont {Carbone}},\ }\href {\doibase
  10.1103/PhysRevLett.96.156401} {\bibfield  {journal} {\bibinfo  {journal}
  {Phys. Rev. Lett.}\ }\textbf {\bibinfo {volume} {96}},\ \bibinfo {pages}
  {156401} (\bibinfo {year} {2006})}\BibitemShut {NoStop}%
\bibitem [{\citenamefont {Kraus}\ and\ \citenamefont
  {Zilberberg}(2012)}]{Kraus_Zilberberg_Topological_PRB}%
  \BibitemOpen
  \bibfield  {author} {\bibinfo {author} {\bibfnamefont {Y.~E.}\ \bibnamefont
  {Kraus}}\ and\ \bibinfo {author} {\bibfnamefont {O.}~\bibnamefont
  {Zilberberg}},\ }\href {\doibase 10.1103/PhysRevLett.109.116404} {\bibfield
  {journal} {\bibinfo  {journal} {Phys. Rev. Lett.}\ }\textbf {\bibinfo
  {volume} {109}},\ \bibinfo {pages} {116404} (\bibinfo {year}
  {2012})}\BibitemShut {NoStop}%
\bibitem [{\citenamefont {Kraus}\ \emph
  {et~al.}(2012{\natexlab{a}})\citenamefont {Kraus}, \citenamefont {Lahini},
  \citenamefont {Ringel}, \citenamefont {Verbin},\ and\ \citenamefont
  {Zilberberg}}]{ZilberbergQC}%
  \BibitemOpen
  \bibfield  {author} {\bibinfo {author} {\bibfnamefont {Y.~E.}\ \bibnamefont
  {Kraus}}, \bibinfo {author} {\bibfnamefont {Y.}~\bibnamefont {Lahini}},
  \bibinfo {author} {\bibfnamefont {Z.}~\bibnamefont {Ringel}}, \bibinfo
  {author} {\bibfnamefont {M.}~\bibnamefont {Verbin}}, \ and\ \bibinfo {author}
  {\bibfnamefont {O.}~\bibnamefont {Zilberberg}},\ }\href {\doibase
  10.1103/PhysRevLett.109.106402} {\bibfield  {journal} {\bibinfo  {journal}
  {Phys. Rev. Lett.}\ }\textbf {\bibinfo {volume} {109}},\ \bibinfo {pages}
  {106402} (\bibinfo {year} {2012}{\natexlab{a}})}\BibitemShut {NoStop}%
\bibitem [{\citenamefont {Mei}\ \emph {et~al.}(2012)\citenamefont {Mei},
  \citenamefont {Zhu}, \citenamefont {Zhang}, \citenamefont {Oh},\ and\
  \citenamefont {Goldman}}]{boundaryphenomena2}%
  \BibitemOpen
  \bibfield  {author} {\bibinfo {author} {\bibfnamefont {F.}~\bibnamefont
  {Mei}}, \bibinfo {author} {\bibfnamefont {S.-L.}\ \bibnamefont {Zhu}},
  \bibinfo {author} {\bibfnamefont {Z.-M.}\ \bibnamefont {Zhang}}, \bibinfo
  {author} {\bibfnamefont {C.~H.}\ \bibnamefont {Oh}}, \ and\ \bibinfo {author}
  {\bibfnamefont {N.}~\bibnamefont {Goldman}},\ }\href {\doibase
  10.1103/PhysRevA.85.013638} {\bibfield  {journal} {\bibinfo  {journal} {Phys.
  Rev. A}\ }\textbf {\bibinfo {volume} {85}},\ \bibinfo {pages} {013638}
  (\bibinfo {year} {2012})}\BibitemShut {NoStop}%
\bibitem [{\citenamefont {Verbin}\ \emph {et~al.}(2013)\citenamefont {Verbin},
  \citenamefont {Zilberberg}, \citenamefont {Kraus}, \citenamefont {Lahini},\
  and\ \citenamefont {Silberberg}}]{verbin-photonic-topological_PRL}%
  \BibitemOpen
  \bibfield  {author} {\bibinfo {author} {\bibfnamefont {M.}~\bibnamefont
  {Verbin}}, \bibinfo {author} {\bibfnamefont {O.}~\bibnamefont {Zilberberg}},
  \bibinfo {author} {\bibfnamefont {Y.~E.}\ \bibnamefont {Kraus}}, \bibinfo
  {author} {\bibfnamefont {Y.}~\bibnamefont {Lahini}}, \ and\ \bibinfo {author}
  {\bibfnamefont {Y.}~\bibnamefont {Silberberg}},\ }\href {\doibase
  10.1103/PhysRevLett.110.076403} {\bibfield  {journal} {\bibinfo  {journal}
  {Phys. Rev. Lett.}\ }\textbf {\bibinfo {volume} {110}},\ \bibinfo {pages}
  {076403} (\bibinfo {year} {2013})}\BibitemShut {NoStop}%
\bibitem [{\citenamefont {Madsen}\ \emph {et~al.}(2013)\citenamefont {Madsen},
  \citenamefont {Bergholtz},\ and\ \citenamefont {Brouwer}}]{brouwerpaper}%
  \BibitemOpen
  \bibfield  {author} {\bibinfo {author} {\bibfnamefont {K.~A.}\ \bibnamefont
  {Madsen}}, \bibinfo {author} {\bibfnamefont {E.~J.}\ \bibnamefont
  {Bergholtz}}, \ and\ \bibinfo {author} {\bibfnamefont {P.~W.}\ \bibnamefont
  {Brouwer}},\ }\href {\doibase 10.1103/PhysRevB.88.125118} {\bibfield
  {journal} {\bibinfo  {journal} {Phys. Rev. B}\ }\textbf {\bibinfo {volume}
  {88}},\ \bibinfo {pages} {125118} (\bibinfo {year} {2013})}\BibitemShut
  {NoStop}%
\bibitem [{\citenamefont {Goldman}\ \emph {et~al.}(2013)\citenamefont
  {Goldman}, \citenamefont {Kong}, \citenamefont {Kreyssig}, \citenamefont
  {Jesche}, \citenamefont {Ramazanoglu}, \citenamefont {Dennis}, \citenamefont
  {Bud'ko},\ and\ \citenamefont {Canfield}}]{ames-QCs}%
  \BibitemOpen
  \bibfield  {author} {\bibinfo {author} {\bibfnamefont {A.~I.}\ \bibnamefont
  {Goldman}}, \bibinfo {author} {\bibfnamefont {T.}~\bibnamefont {Kong}},
  \bibinfo {author} {\bibfnamefont {A.}~\bibnamefont {Kreyssig}}, \bibinfo
  {author} {\bibfnamefont {A.}~\bibnamefont {Jesche}}, \bibinfo {author}
  {\bibfnamefont {M.}~\bibnamefont {Ramazanoglu}}, \bibinfo {author}
  {\bibfnamefont {K.~W.}\ \bibnamefont {Dennis}}, \bibinfo {author}
  {\bibfnamefont {S.~L.}\ \bibnamefont {Bud'ko}}, \ and\ \bibinfo {author}
  {\bibfnamefont {P.~C.}\ \bibnamefont {Canfield}},\ }\href
  {http://dx.doi.org/10.1038/nmat3672} {\bibfield  {journal} {\bibinfo
  {journal} {Nature Materials}\ }\textbf {\bibinfo {volume} {12}},\ \bibinfo
  {pages} {714} (\bibinfo {year} {2013})}\BibitemShut {NoStop}%
\bibitem [{\citenamefont {Lang}\ \emph {et~al.}(2012)\citenamefont {Lang},
  \citenamefont {Cai},\ and\ \citenamefont
  {Chen}}]{hofstadter-superlattice-coldatom-proposal}%
  \BibitemOpen
  \bibfield  {author} {\bibinfo {author} {\bibfnamefont {L.-J.}\ \bibnamefont
  {Lang}}, \bibinfo {author} {\bibfnamefont {X.}~\bibnamefont {Cai}}, \ and\
  \bibinfo {author} {\bibfnamefont {S.}~\bibnamefont {Chen}},\ }\href {\doibase
  10.1103/PhysRevLett.108.220401} {\bibfield  {journal} {\bibinfo  {journal}
  {Phys. Rev. Lett.}\ }\textbf {\bibinfo {volume} {108}},\ \bibinfo {pages}
  {220401} (\bibinfo {year} {2012})}\BibitemShut {NoStop}%
\bibitem [{\citenamefont {Lye}\ \emph {et~al.}(2007)\citenamefont {Lye},
  \citenamefont {Fallani}, \citenamefont {Fort}, \citenamefont {Guarrera},
  \citenamefont {Modugno}, \citenamefont {Wiersma},\ and\ \citenamefont
  {Inguscio}}]{Lye-Inguscio-interaction-localization_PRA}%
  \BibitemOpen
  \bibfield  {author} {\bibinfo {author} {\bibfnamefont {J.~E.}\ \bibnamefont
  {Lye}}, \bibinfo {author} {\bibfnamefont {L.}~\bibnamefont {Fallani}},
  \bibinfo {author} {\bibfnamefont {C.}~\bibnamefont {Fort}}, \bibinfo {author}
  {\bibfnamefont {V.}~\bibnamefont {Guarrera}}, \bibinfo {author}
  {\bibfnamefont {M.}~\bibnamefont {Modugno}}, \bibinfo {author} {\bibfnamefont
  {D.~S.}\ \bibnamefont {Wiersma}}, \ and\ \bibinfo {author} {\bibfnamefont
  {M.}~\bibnamefont {Inguscio}},\ }\href {\doibase 10.1103/PhysRevA.75.061603}
  {\bibfield  {journal} {\bibinfo  {journal} {Phys. Rev. A}\ }\textbf {\bibinfo
  {volume} {75}},\ \bibinfo {pages} {061603} (\bibinfo {year}
  {2007})}\BibitemShut {NoStop}%
\bibitem [{\citenamefont {Roati}\ \emph {et~al.}(2008)\citenamefont {Roati},
  \citenamefont {D'Errico}, \citenamefont {Fallani}, \citenamefont {Fattori},
  \citenamefont {Fort}, \citenamefont {Zaccanti}, \citenamefont {Modugno},
  \citenamefont {Modugno},\ and\ \citenamefont
  {Inguscio}}]{inguscio-andersonloc}%
  \BibitemOpen
  \bibfield  {author} {\bibinfo {author} {\bibfnamefont {G.}~\bibnamefont
  {Roati}}, \bibinfo {author} {\bibfnamefont {C.}~\bibnamefont {D'Errico}},
  \bibinfo {author} {\bibfnamefont {L.}~\bibnamefont {Fallani}}, \bibinfo
  {author} {\bibfnamefont {M.}~\bibnamefont {Fattori}}, \bibinfo {author}
  {\bibfnamefont {C.}~\bibnamefont {Fort}}, \bibinfo {author} {\bibfnamefont
  {M.}~\bibnamefont {Zaccanti}}, \bibinfo {author} {\bibfnamefont
  {G.}~\bibnamefont {Modugno}}, \bibinfo {author} {\bibfnamefont
  {M.}~\bibnamefont {Modugno}}, \ and\ \bibinfo {author} {\bibfnamefont
  {M.}~\bibnamefont {Inguscio}},\ }\href
  {http://dx.doi.org/10.1038/nature07071} {\bibfield  {journal} {\bibinfo
  {journal} {Nature}\ }\textbf {\bibinfo {volume} {453}},\ \bibinfo {pages}
  {895} (\bibinfo {year} {2008})}\BibitemShut {NoStop}%
\bibitem [{\citenamefont {{Schreiber}}\ \emph {et~al.}(2015)\citenamefont
  {{Schreiber}}, \citenamefont {{Hodgman}}, \citenamefont {{Bordia}},
  \citenamefont {{L{\"u}schen}}, \citenamefont {{Fischer}}, \citenamefont
  {{Vosk}}, \citenamefont {{Altman}}, \citenamefont {{Schneider}},\ and\
  \citenamefont {{Bloch}}}]{bloch-mbl}%
  \BibitemOpen
  \bibfield  {author} {\bibinfo {author} {\bibfnamefont {M.}~\bibnamefont
  {{Schreiber}}}, \bibinfo {author} {\bibfnamefont {S.~S.}\ \bibnamefont
  {{Hodgman}}}, \bibinfo {author} {\bibfnamefont {P.}~\bibnamefont {{Bordia}}},
  \bibinfo {author} {\bibfnamefont {H.~P.}\ \bibnamefont {{L{\"u}schen}}},
  \bibinfo {author} {\bibfnamefont {M.~H.}\ \bibnamefont {{Fischer}}}, \bibinfo
  {author} {\bibfnamefont {R.}~\bibnamefont {{Vosk}}}, \bibinfo {author}
  {\bibfnamefont {E.}~\bibnamefont {{Altman}}}, \bibinfo {author}
  {\bibfnamefont {U.}~\bibnamefont {{Schneider}}}, \ and\ \bibinfo {author}
  {\bibfnamefont {I.}~\bibnamefont {{Bloch}}},\ }\href@noop {} {\bibfield
  {journal} {\bibinfo  {journal} {arXiv:1501.05661}\ } (\bibinfo {year}
  {2015})}\BibitemShut {NoStop}%
\bibitem [{\citenamefont {Guidoni}\ \emph {et~al.}(1997)\citenamefont
  {Guidoni}, \citenamefont {Trich\'e}, \citenamefont {Verkerk},\ and\
  \citenamefont {Grynberg}}]{Verkerk-Grynberg-5fold_PRL}%
  \BibitemOpen
  \bibfield  {author} {\bibinfo {author} {\bibfnamefont {L.}~\bibnamefont
  {Guidoni}}, \bibinfo {author} {\bibfnamefont {C.}~\bibnamefont {Trich\'e}},
  \bibinfo {author} {\bibfnamefont {P.}~\bibnamefont {Verkerk}}, \ and\
  \bibinfo {author} {\bibfnamefont {G.}~\bibnamefont {Grynberg}},\ }\href
  {\doibase 10.1103/PhysRevLett.79.3363} {\bibfield  {journal} {\bibinfo
  {journal} {Phys. Rev. Lett.}\ }\textbf {\bibinfo {volume} {79}},\ \bibinfo
  {pages} {3363} (\bibinfo {year} {1997})}\BibitemShut {NoStop}%
\bibitem [{\citenamefont {Guidoni}\ \emph {et~al.}(1999)\citenamefont
  {Guidoni}, \citenamefont {D\'epret}, \citenamefont {di~Stefano},\ and\
  \citenamefont {Verkerk}}]{Verkerk-5fold-diffusion_PRA}%
  \BibitemOpen
  \bibfield  {author} {\bibinfo {author} {\bibfnamefont {L.}~\bibnamefont
  {Guidoni}}, \bibinfo {author} {\bibfnamefont {B.}~\bibnamefont {D\'epret}},
  \bibinfo {author} {\bibfnamefont {A.}~\bibnamefont {di~Stefano}}, \ and\
  \bibinfo {author} {\bibfnamefont {P.}~\bibnamefont {Verkerk}},\ }\href
  {\doibase 10.1103/PhysRevA.60.R4233} {\bibfield  {journal} {\bibinfo
  {journal} {Phys. Rev. A}\ }\textbf {\bibinfo {volume} {60}},\ \bibinfo
  {pages} {R4233} (\bibinfo {year} {1999})}\BibitemShut {NoStop}%
\bibitem [{\citenamefont {Corcovilos}\ and\ \citenamefont
  {Mittal}(2019)}]{QC-corcovilos19}%
  \BibitemOpen
  \bibfield  {author} {\bibinfo {author} {\bibfnamefont {T.~A.}\ \bibnamefont
  {Corcovilos}}\ and\ \bibinfo {author} {\bibfnamefont {J.}~\bibnamefont
  {Mittal}},\ }\href {\doibase 10.1364/AO.58.002256} {\bibfield  {journal}
  {\bibinfo  {journal} {Appl. Opt.}\ }\textbf {\bibinfo {volume} {58}},\
  \bibinfo {pages} {2256} (\bibinfo {year} {2019})}\BibitemShut {NoStop}%
\bibitem [{\citenamefont {Singh}\ \emph {et~al.}(2015)\citenamefont {Singh},
  \citenamefont {Saha}, \citenamefont {Parameswaran},\ and\ \citenamefont
  {Weld}}]{fibonacci-PRA}%
  \BibitemOpen
  \bibfield  {author} {\bibinfo {author} {\bibfnamefont {K.}~\bibnamefont
  {Singh}}, \bibinfo {author} {\bibfnamefont {K.}~\bibnamefont {Saha}},
  \bibinfo {author} {\bibfnamefont {S.~A.}\ \bibnamefont {Parameswaran}}, \
  and\ \bibinfo {author} {\bibfnamefont {D.~M.}\ \bibnamefont {Weld}},\ }\href
  {\doibase 10.1103/PhysRevA.92.063426} {\bibfield  {journal} {\bibinfo
  {journal} {Phys. Rev. A}\ }\textbf {\bibinfo {volume} {92}},\ \bibinfo
  {pages} {063426} (\bibinfo {year} {2015})}\BibitemShut {NoStop}%
\bibitem [{\citenamefont {Gadway}\ \emph {et~al.}(2013)\citenamefont {Gadway},
  \citenamefont {Reeves}, \citenamefont {Krinner},\ and\ \citenamefont
  {Schneble}}]{Gadway-schneble_PRL}%
  \BibitemOpen
  \bibfield  {author} {\bibinfo {author} {\bibfnamefont {B.}~\bibnamefont
  {Gadway}}, \bibinfo {author} {\bibfnamefont {J.}~\bibnamefont {Reeves}},
  \bibinfo {author} {\bibfnamefont {L.}~\bibnamefont {Krinner}}, \ and\
  \bibinfo {author} {\bibfnamefont {D.}~\bibnamefont {Schneble}},\ }\href
  {\doibase 10.1103/PhysRevLett.110.190401} {\bibfield  {journal} {\bibinfo
  {journal} {Phys. Rev. Lett.}\ }\textbf {\bibinfo {volume} {110}},\ \bibinfo
  {pages} {190401} (\bibinfo {year} {2013})}\BibitemShut {NoStop}%
\bibitem [{\citenamefont {Viebahn}\ \emph {et~al.}(2019)\citenamefont
  {Viebahn}, \citenamefont {Sbroscia}, \citenamefont {Carter}, \citenamefont
  {Yu},\ and\ \citenamefont {Schneider}}]{Viebahn-Schneider-8fold_PRL}%
  \BibitemOpen
  \bibfield  {author} {\bibinfo {author} {\bibfnamefont {K.}~\bibnamefont
  {Viebahn}}, \bibinfo {author} {\bibfnamefont {M.}~\bibnamefont {Sbroscia}},
  \bibinfo {author} {\bibfnamefont {E.}~\bibnamefont {Carter}}, \bibinfo
  {author} {\bibfnamefont {J.-C.}\ \bibnamefont {Yu}}, \ and\ \bibinfo {author}
  {\bibfnamefont {U.}~\bibnamefont {Schneider}},\ }\href {\doibase
  10.1103/PhysRevLett.122.110404} {\bibfield  {journal} {\bibinfo  {journal}
  {Phys. Rev. Lett.}\ }\textbf {\bibinfo {volume} {122}},\ \bibinfo {pages}
  {110404} (\bibinfo {year} {2019})}\BibitemShut {NoStop}%
\bibitem [{\citenamefont {Johnstone}\ \emph {et~al.}(2019)\citenamefont
  {Johnstone}, \citenamefont {Ohberg},\ and\ \citenamefont
  {Duncan}}]{8fold-phase-transition}%
  \BibitemOpen
  \bibfield  {author} {\bibinfo {author} {\bibfnamefont {D.}~\bibnamefont
  {Johnstone}}, \bibinfo {author} {\bibfnamefont {P.}~\bibnamefont {Ohberg}}, \
  and\ \bibinfo {author} {\bibfnamefont {C.~W.}\ \bibnamefont {Duncan}},\
  }\href@noop {} {\bibfield  {journal} {\bibinfo  {journal} {arXiv:1904.12870}\
  } (\bibinfo {year} {2019})}\BibitemShut {NoStop}%
\bibitem [{\citenamefont {Kromer}\ \emph {et~al.}(2012)\citenamefont {Kromer},
  \citenamefont {Schmiedeberg}, \citenamefont {Roth},\ and\ \citenamefont
  {Stark}}]{Kromer-PhasonTrajectories_PRL}%
  \BibitemOpen
  \bibfield  {author} {\bibinfo {author} {\bibfnamefont {J.~A.}\ \bibnamefont
  {Kromer}}, \bibinfo {author} {\bibfnamefont {M.}~\bibnamefont
  {Schmiedeberg}}, \bibinfo {author} {\bibfnamefont {J.}~\bibnamefont {Roth}},
  \ and\ \bibinfo {author} {\bibfnamefont {H.}~\bibnamefont {Stark}},\ }\href
  {\doibase 10.1103/PhysRevLett.108.218301} {\bibfield  {journal} {\bibinfo
  {journal} {Phys. Rev. Lett.}\ }\textbf {\bibinfo {volume} {108}},\ \bibinfo
  {pages} {218301} (\bibinfo {year} {2012})}\BibitemShut {NoStop}%
\bibitem [{\citenamefont {Widom}(2008)}]{Widom_PhasonReview}%
  \BibitemOpen
  \bibfield  {author} {\bibinfo {author} {\bibfnamefont {M.}~\bibnamefont
  {Widom}},\ }\href@noop {} {\bibfield  {journal} {\bibinfo  {journal}
  {Philosophical Magazine}\ }\textbf {\bibinfo {volume} {88}},\ \bibinfo
  {pages} {2339} (\bibinfo {year} {2008})}\BibitemShut {NoStop}%
\bibitem [{\citenamefont {Aubry}\ and\ \citenamefont
  {Andr{\'e}}(1980)}]{aubryandre}%
  \BibitemOpen
  \bibfield  {author} {\bibinfo {author} {\bibfnamefont {S.}~\bibnamefont
  {Aubry}}\ and\ \bibinfo {author} {\bibfnamefont {G.}~\bibnamefont
  {Andr{\'e}}},\ }\href@noop {} {\bibfield  {journal} {\bibinfo  {journal}
  {Ann. Isr. Phys. Soc.}\ }\textbf {\bibinfo {volume} {3}},\ \bibinfo {pages}
  {133} (\bibinfo {year} {1980})}\BibitemShut {NoStop}%
\bibitem [{\citenamefont {Harper}(1955)}]{harpermodel}%
  \BibitemOpen
  \bibfield  {author} {\bibinfo {author} {\bibfnamefont {P.~G.}\ \bibnamefont
  {Harper}},\ }\href {http://stacks.iop.org/0370-1298/68/i=10/a=304} {\bibfield
   {journal} {\bibinfo  {journal} {Proceedings of the Physical Society. Section
  A}\ }\textbf {\bibinfo {volume} {68}},\ \bibinfo {pages} {874} (\bibinfo
  {year} {1955})}\BibitemShut {NoStop}%
\bibitem [{\citenamefont {Boers}\ \emph {et~al.}(2007)\citenamefont {Boers},
  \citenamefont {Goedeke}, \citenamefont {Hinrichs},\ and\ \citenamefont
  {Holthaus}}]{spme-boers_PRA07}%
  \BibitemOpen
  \bibfield  {author} {\bibinfo {author} {\bibfnamefont {D.~J.}\ \bibnamefont
  {Boers}}, \bibinfo {author} {\bibfnamefont {B.}~\bibnamefont {Goedeke}},
  \bibinfo {author} {\bibfnamefont {D.}~\bibnamefont {Hinrichs}}, \ and\
  \bibinfo {author} {\bibfnamefont {M.}~\bibnamefont {Holthaus}},\ }\href
  {\doibase 10.1103/PhysRevA.75.063404} {\bibfield  {journal} {\bibinfo
  {journal} {Phys. Rev. A}\ }\textbf {\bibinfo {volume} {75}},\ \bibinfo
  {pages} {063404} (\bibinfo {year} {2007})}\BibitemShut {NoStop}%
\bibitem [{\citenamefont {Li}\ \emph {et~al.}(2017)\citenamefont {Li},
  \citenamefont {Li},\ and\ \citenamefont {Das~Sarma}}]{spme-das-sarma_PRB}%
  \BibitemOpen
  \bibfield  {author} {\bibinfo {author} {\bibfnamefont {X.}~\bibnamefont
  {Li}}, \bibinfo {author} {\bibfnamefont {X.}~\bibnamefont {Li}}, \ and\
  \bibinfo {author} {\bibfnamefont {S.}~\bibnamefont {Das~Sarma}},\ }\href
  {\doibase 10.1103/PhysRevB.96.085119} {\bibfield  {journal} {\bibinfo
  {journal} {Phys. Rev. B}\ }\textbf {\bibinfo {volume} {96}},\ \bibinfo
  {pages} {085119} (\bibinfo {year} {2017})}\BibitemShut {NoStop}%
\bibitem [{\citenamefont {L\"uschen}\ \emph {et~al.}(2018)\citenamefont
  {L\"uschen}, \citenamefont {Scherg}, \citenamefont {Kohlert}, \citenamefont
  {Schreiber}, \citenamefont {Bordia}, \citenamefont {Li}, \citenamefont
  {Das~Sarma},\ and\ \citenamefont {Bloch}}]{spme-bloch_PRL18}%
  \BibitemOpen
  \bibfield  {author} {\bibinfo {author} {\bibfnamefont {H.~P.}\ \bibnamefont
  {L\"uschen}}, \bibinfo {author} {\bibfnamefont {S.}~\bibnamefont {Scherg}},
  \bibinfo {author} {\bibfnamefont {T.}~\bibnamefont {Kohlert}}, \bibinfo
  {author} {\bibfnamefont {M.}~\bibnamefont {Schreiber}}, \bibinfo {author}
  {\bibfnamefont {P.}~\bibnamefont {Bordia}}, \bibinfo {author} {\bibfnamefont
  {X.}~\bibnamefont {Li}}, \bibinfo {author} {\bibfnamefont {S.}~\bibnamefont
  {Das~Sarma}}, \ and\ \bibinfo {author} {\bibfnamefont {I.}~\bibnamefont
  {Bloch}},\ }\href {\doibase 10.1103/PhysRevLett.120.160404} {\bibfield
  {journal} {\bibinfo  {journal} {Phys. Rev. Lett.}\ }\textbf {\bibinfo
  {volume} {120}},\ \bibinfo {pages} {160404} (\bibinfo {year}
  {2018})}\BibitemShut {NoStop}%
\bibitem [{\citenamefont {Kohlert}\ \emph {et~al.}(2019)\citenamefont
  {Kohlert}, \citenamefont {Scherg}, \citenamefont {Li}, \citenamefont
  {L\"uschen}, \citenamefont {Das~Sarma}, \citenamefont {Bloch},\ and\
  \citenamefont {Aidelsburger}}]{monika-mbl-spme}%
  \BibitemOpen
  \bibfield  {author} {\bibinfo {author} {\bibfnamefont {T.}~\bibnamefont
  {Kohlert}}, \bibinfo {author} {\bibfnamefont {S.}~\bibnamefont {Scherg}},
  \bibinfo {author} {\bibfnamefont {X.}~\bibnamefont {Li}}, \bibinfo {author}
  {\bibfnamefont {H.~P.}\ \bibnamefont {L\"uschen}}, \bibinfo {author}
  {\bibfnamefont {S.}~\bibnamefont {Das~Sarma}}, \bibinfo {author}
  {\bibfnamefont {I.}~\bibnamefont {Bloch}}, \ and\ \bibinfo {author}
  {\bibfnamefont {M.}~\bibnamefont {Aidelsburger}},\ }\href {\doibase
  10.1103/PhysRevLett.122.170403} {\bibfield  {journal} {\bibinfo  {journal}
  {Phys. Rev. Lett.}\ }\textbf {\bibinfo {volume} {122}},\ \bibinfo {pages}
  {170403} (\bibinfo {year} {2019})}\BibitemShut {NoStop}%
\bibitem [{\citenamefont {Eckardt}(2017)}]{Eckardt2017}%
  \BibitemOpen
  \bibfield  {author} {\bibinfo {author} {\bibfnamefont {A.}~\bibnamefont
  {Eckardt}},\ }\href {\doibase 10.1103/RevModPhys.89.011004} {\bibfield
  {journal} {\bibinfo  {journal} {Rev. Mod. Phys.}\ }\textbf {\bibinfo {volume}
  {89}},\ \bibinfo {pages} {011004} (\bibinfo {year} {2017})}\BibitemShut
  {NoStop}%
\bibitem [{\citenamefont {Weinberg}\ \emph {et~al.}(2015)\citenamefont
  {Weinberg}, \citenamefont {\"Olschl\"ager}, \citenamefont {Str\"ater},
  \citenamefont {Prelle}, \citenamefont {Eckardt}, \citenamefont {Sengstock},\
  and\ \citenamefont {Simonet}}]{interband_eckardt_sengstock_simonet}%
  \BibitemOpen
  \bibfield  {author} {\bibinfo {author} {\bibfnamefont {M.}~\bibnamefont
  {Weinberg}}, \bibinfo {author} {\bibfnamefont {C.}~\bibnamefont
  {\"Olschl\"ager}}, \bibinfo {author} {\bibfnamefont {C.}~\bibnamefont
  {Str\"ater}}, \bibinfo {author} {\bibfnamefont {S.}~\bibnamefont {Prelle}},
  \bibinfo {author} {\bibfnamefont {A.}~\bibnamefont {Eckardt}}, \bibinfo
  {author} {\bibfnamefont {K.}~\bibnamefont {Sengstock}}, \ and\ \bibinfo
  {author} {\bibfnamefont {J.}~\bibnamefont {Simonet}},\ }\href {\doibase
  10.1103/PhysRevA.92.043621} {\bibfield  {journal} {\bibinfo  {journal} {Phys.
  Rev. A}\ }\textbf {\bibinfo {volume} {92}},\ \bibinfo {pages} {043621}
  (\bibinfo {year} {2015})}\BibitemShut {NoStop}%
\bibitem [{\citenamefont {Reitter}\ \emph {et~al.}(2017)\citenamefont
  {Reitter}, \citenamefont {N\"ager}, \citenamefont {Wintersperger},
  \citenamefont {Str\"ater}, \citenamefont {Bloch}, \citenamefont {Eckardt},\
  and\ \citenamefont {Schneider}}]{latticeheating-blochgroup}%
  \BibitemOpen
  \bibfield  {author} {\bibinfo {author} {\bibfnamefont {M.}~\bibnamefont
  {Reitter}}, \bibinfo {author} {\bibfnamefont {J.}~\bibnamefont {N\"ager}},
  \bibinfo {author} {\bibfnamefont {K.}~\bibnamefont {Wintersperger}}, \bibinfo
  {author} {\bibfnamefont {C.}~\bibnamefont {Str\"ater}}, \bibinfo {author}
  {\bibfnamefont {I.}~\bibnamefont {Bloch}}, \bibinfo {author} {\bibfnamefont
  {A.}~\bibnamefont {Eckardt}}, \ and\ \bibinfo {author} {\bibfnamefont
  {U.}~\bibnamefont {Schneider}},\ }\href {\doibase
  10.1103/PhysRevLett.119.200402} {\bibfield  {journal} {\bibinfo  {journal}
  {Phys. Rev. Lett.}\ }\textbf {\bibinfo {volume} {119}},\ \bibinfo {pages}
  {200402} (\bibinfo {year} {2017})}\BibitemShut {NoStop}%
\bibitem [{\citenamefont {McKay}\ \emph {et~al.}(2009)\citenamefont {McKay},
  \citenamefont {White},\ and\ \citenamefont {DeMarco}}]{demarco-bandmap}%
  \BibitemOpen
  \bibfield  {author} {\bibinfo {author} {\bibfnamefont {D.}~\bibnamefont
  {McKay}}, \bibinfo {author} {\bibfnamefont {M.}~\bibnamefont {White}}, \ and\
  \bibinfo {author} {\bibfnamefont {B.}~\bibnamefont {DeMarco}},\ }\href
  {\doibase 10.1103/PhysRevA.79.063605} {\bibfield  {journal} {\bibinfo
  {journal} {Phys. Rev. A}\ }\textbf {\bibinfo {volume} {79}},\ \bibinfo
  {pages} {063605} (\bibinfo {year} {2009})}\BibitemShut {NoStop}%
\bibitem [{Sup()}]{SuppMat}%
  \BibitemOpen
  \href@noop {} {}\bibinfo {note} {See Supplemental Material for a discussion
  on numerical calculations and an analytic explanation for high-order phasonic
  multiphoton resonances.}\BibitemShut {Stop}%
\bibitem [{\citenamefont {Str\"ater}\ and\ \citenamefont
  {Eckardt}(2016)}]{Straeter2016}%
  \BibitemOpen
  \bibfield  {author} {\bibinfo {author} {\bibfnamefont {C.}~\bibnamefont
  {Str\"ater}}\ and\ \bibinfo {author} {\bibfnamefont {A.}~\bibnamefont
  {Eckardt}},\ }\href {\doibase 10.1515/zna-2016-0129} {\bibfield  {journal}
  {\bibinfo  {journal} {Z.\ Naturforsch.\ A}\ }\textbf {\bibinfo {volume}
  {71}},\ \bibinfo {pages} {909} (\bibinfo {year} {2016})}\BibitemShut
  {NoStop}%
\bibitem [{\citenamefont {Limas}\ \emph {et~al.}(2005)\citenamefont {Limas},
  \citenamefont {Naumis}, \citenamefont {Salazar},\ and\ \citenamefont
  {Wang}}]{physletta-phonon-limas05}%
  \BibitemOpen
  \bibfield  {author} {\bibinfo {author} {\bibfnamefont {I.}~\bibnamefont
  {Limas}}, \bibinfo {author} {\bibfnamefont {G.}~\bibnamefont {Naumis}},
  \bibinfo {author} {\bibfnamefont {F.}~\bibnamefont {Salazar}}, \ and\
  \bibinfo {author} {\bibfnamefont {C.}~\bibnamefont {Wang}},\ }\href {\doibase
  https://doi.org/10.1016/j.physleta.2005.01.054} {\bibfield  {journal}
  {\bibinfo  {journal} {Physics Letters A}\ }\textbf {\bibinfo {volume}
  {337}},\ \bibinfo {pages} {141 } (\bibinfo {year} {2005})}\BibitemShut
  {NoStop}%
\bibitem [{\citenamefont {Zhu}\ \emph {et~al.}(1997)\citenamefont {Zhu},
  \citenamefont {Zhu},\ and\ \citenamefont {Ming}}]{science-hhgqc-zhu97}%
  \BibitemOpen
  \bibfield  {author} {\bibinfo {author} {\bibfnamefont {S.-n.}\ \bibnamefont
  {Zhu}}, \bibinfo {author} {\bibfnamefont {Y.-y.}\ \bibnamefont {Zhu}}, \ and\
  \bibinfo {author} {\bibfnamefont {N.-b.}\ \bibnamefont {Ming}},\ }\href
  {\doibase 10.1126/science.278.5339.843} {\bibfield  {journal} {\bibinfo
  {journal} {Science}\ }\textbf {\bibinfo {volume} {278}},\ \bibinfo {pages}
  {843} (\bibinfo {year} {1997})}\BibitemShut {NoStop}%
\bibitem [{\citenamefont {Hofstadter}(1976)}]{hofstadter-theoriginal}%
  \BibitemOpen
  \bibfield  {author} {\bibinfo {author} {\bibfnamefont {D.~R.}\ \bibnamefont
  {Hofstadter}},\ }\href {\doibase 10.1103/PhysRevB.14.2239} {\bibfield
  {journal} {\bibinfo  {journal} {Phys. Rev. B}\ }\textbf {\bibinfo {volume}
  {14}},\ \bibinfo {pages} {2239} (\bibinfo {year} {1976})}\BibitemShut
  {NoStop}%
\bibitem [{\citenamefont {Lindner}\ \emph {et~al.}(2017)\citenamefont
  {Lindner}, \citenamefont {Berg},\ and\ \citenamefont
  {Rudner}}]{Lindner-Rudner-thouless_PRX}%
  \BibitemOpen
  \bibfield  {author} {\bibinfo {author} {\bibfnamefont {N.~H.}\ \bibnamefont
  {Lindner}}, \bibinfo {author} {\bibfnamefont {E.}~\bibnamefont {Berg}}, \
  and\ \bibinfo {author} {\bibfnamefont {M.~S.}\ \bibnamefont {Rudner}},\
  }\href {\doibase 10.1103/PhysRevX.7.011018} {\bibfield  {journal} {\bibinfo
  {journal} {Phys. Rev. X}\ }\textbf {\bibinfo {volume} {7}},\ \bibinfo {pages}
  {011018} (\bibinfo {year} {2017})}\BibitemShut {NoStop}%
\bibitem [{\citenamefont {Kraus}\ \emph
  {et~al.}(2012{\natexlab{b}})\citenamefont {Kraus}, \citenamefont {Lahini},
  \citenamefont {Ringel}, \citenamefont {Verbin},\ and\ \citenamefont
  {Zilberberg}}]{topological-quasicrystals-PRL}%
  \BibitemOpen
  \bibfield  {author} {\bibinfo {author} {\bibfnamefont {Y.~E.}\ \bibnamefont
  {Kraus}}, \bibinfo {author} {\bibfnamefont {Y.}~\bibnamefont {Lahini}},
  \bibinfo {author} {\bibfnamefont {Z.}~\bibnamefont {Ringel}}, \bibinfo
  {author} {\bibfnamefont {M.}~\bibnamefont {Verbin}}, \ and\ \bibinfo {author}
  {\bibfnamefont {O.}~\bibnamefont {Zilberberg}},\ }\href {\doibase
  10.1103/PhysRevLett.109.106402} {\bibfield  {journal} {\bibinfo  {journal}
  {Phys. Rev. Lett.}\ }\textbf {\bibinfo {volume} {109}},\ \bibinfo {pages}
  {106402} (\bibinfo {year} {2012}{\natexlab{b}})}\BibitemShut {NoStop}%
\end{thebibliography}

\bibliographystyle{apsrev4-1}

\clearpage

%%%%%%%%%% Merge with supplemental materials %%%%%%%%%%
\pagebreak
\widetext
\begin{center}
\textbf{\large Supplemental Material for `Phasonic Spectroscopy of a Quantum Gas in a Quasicrystalline Lattice'}
\end{center}
%%%%%%%%%% Merge with supplemental materials %%%%%%%%%%
%%%%%%%%%% Prefix a "S" to all equations, figures, tables and reset the counter %%%%%%%%%%
\setcounter{equation}{0}
\setcounter{figure}{0}
\setcounter{table}{0}
\setcounter{page}{1}
\makeatletter
\renewcommand{\theequation}{S\arabic{equation}}
\renewcommand{\thefigure}{S\arabic{figure}}
\renewcommand{\bibnumfmt}[1]{[S#1]}
\renewcommand{\citenumfont}[1]{S#1}
%%%%%%%%%% Prefix a "S" to all equations, figures, tables and reset the counter %%%%%%%%%%

\section{Description of numerical calculations}
\subsection{Phasonic time-evolution simulations}
Here we describe the numerical calculations used to produce the theoretical predictions discussed in the main text of the paper. The simulations shown in Figs.~2f, 3a  are done in the formalism of exact time evolution without interactions using $N = 43$ lattice sites with open boundary conditions, taking the 5 lowest Bloch bands into account. First, we obtain the Wannier functions of the principal $1064\,\mathrm{nm}$ lattice, which are used as the basis. This in turn allows us to calculate coupling matrix elements between different lattice sites and bands. Note that we do not restrict the calculation of the matrix elements to the nearest neighbors, and we include couplings to further sites if relevant. In the phasonic driving regime, the relative displacement of the two lattices is constantly changing during the period of the drive and recalculation of matrix elements at every integration step becomes cumbersome. Therefore we discretize the relative displacements using $M = 100$ points and store the matrix elements calculated for different displacements. During the time evolution, we then interpolate linearly between these points to obtain matrix elements at each precise instant of time. For the integrator, we set the relative and the absolute tolerance to $10^{-10}$, and after the full driving time, which is set to equal the experimental driving time, we check that the norm of the wave function is maintained up to $10^{-6}$. Due to the immense volume of computations needed for a full simulation of an ensemble of initial states, we take a pure state constructed in such a way that the probability of a specific eigenstate is in accordance with a thermal ensemble at $T=30~\mathrm{nK}$. This is done by introducing random phase differences between neighboring sites, the distribution of which is determined by the temperature. However, almost identical results to those depicted in Fig.~2f can be obtained using the ground state as the initial state, which indicates that finite temperature has minimal effect on the dynamics in these simulations. After the modulation, we project the state to the undriven lattice bands and plot the minimum ground band occupation.
%instead of projecting the wavefunction into free space momentum states as is done in the experiment 
%(IN THE EXPERIMENT NO PROJECTION IS DONE, BUT RATHER AN ADIABATIC SWITCH OFF OF THE LATTICE, USUALLY 
%(FOR A SINGLE COLOR LATTICE) CALLED BAND MAPPING, I WOULD REMOVE THE SUBSENTENCE STARING WITH INSTEAD ...), 

\subsection{Theoretical densities of states for comparison with experimental spectra}

\textbf{Data generation for Fig.~4(d) and 4(e).} In Fig.~4, panels (d) and (e) contrast the theoretically computed densities of states of, respectively, noninteracting and interacting BEC. In both cases, having obtained energy eigenvalues $E_i$, we calculate and plot the corresponding transition frequencies $f_i = (E_i-E_0) / h$ between the $i$-th excited state and the ground-state energy $E_0$. For the interacting case in Fig.~4(e), we describe the condensate by the variational wave-function $\psi(x) \, g_\perp(r_\perp;x)$, where $\psi$ describes the longitudinal direction $x$ and $g_\perp$ is a Gaussian wave-function in radial transverse direction $r_\perp$ with an $x$-dependent width. The obtained spectral minigaps are, in general, quite consistent with their experimental counterparts. In particular, the $V_S$-dependent blueshift is reduced when interactions are taken into account. However, a conspicuous excitation that merges into the main resonance line from above is not reproduced.

\textbf{Data generation for Fig.~4(f).}
To account for the indicated discrepancy, we performed additional modeling using the obtained data. In particular, we calculated and plotted frequencies $f_{ij} = (E_i-E_j) / h$ corresponding to transitions between all pairs of eigenstates $i$ and $j$, with the restriction $(E_j-E_0) \leqslant 1.5E_{R,S}$. In other words, we assumed non-adiabatic loading into the localized regime and allowed the transitions to originate not only from the very bottom of the lowest energy band but also from low-lying excited states in the specified energy window. This scenario enabled us to qualitatively reproduce the hitherto missing characteristic excitation seen in the experiment. 
%it also introduced too much broadening than is observed experimentally. We take this as an indication that the `anomalous' excitation may result from interband transitions involving multi-step interaction-mediated processes: in a first step excited states within the lowest band are occupied, and from there higher bands are populated in a second step. 

\textbf{Further details of time-independent GP calculations for the dipolar drive.}
Let us detail the theoretical approach used to assess the role of particle interactions, keeping in mind that the scattering length for $^{84}$Sr atoms is $\tilde{a} = 124\, a_B$, and there are around $80,000$ particles in the experiment. For the sake of feasibility, we stay within the mean-field approximation and solve the Gross-Pitaevskii-type equation to calculate the effective potential seen by the atoms. %Having obtained the effective potential, we calculate energy eigenvalues $E_i$ that correspond to the $5$ lowest Bloch bands for $N = 51$ lattice sites.
%
%In this subsection, we detail the theoretical approach used to obtain the numerical results shown as dots in Figs.~4c and 4d. Here, the idea is to assess the role of particle interactions, keeping in mind that the scattering length for $^{84}$Sr atoms is $\tilde{a} = 124\, a_B$ and there are around 80,000 particles in the experiment. For the sake of feasibility, we stay within the mean-field approximation and solve the Gross-Pitaevskii-type equation to calculate the the effective potential seen by the atoms. Having obtained the effective potential, we calculate energy eigenvalues $E_i$ that correspond to the $5$ lowest Bloch bands for $N = 51$ lattice sites.
%
We take advantage of the cylindrical symmetry of our potential and describe the radial distribution by a Gaussian profile, whose width is a function of the longitudinal coordinate $x$. This dependence on the coordinate allows us to take into account the non-translationally invariant character of the quasiperiodic lattice. Hence the ansatz for the normalized wave function reads
\begin{equation}
  \Psi(r) = \mathcal{N} \psi(x) g_{\perp}(r_{\perp}; x),    
\end{equation}
with the longitudinal wave function $\psi(x)$ and transverse wave function
\begin{equation}
  g_{\perp}(r_{\perp}; x) = \frac{1}{\sqrt{\pi b^2(x)}}\, e^{-r_{\perp}^2/2b^2(x)}.
\end{equation}
From variation under the constraint $\int\! dx |\psi(x)|^2=1$, we obtain  Gross-Pitaevskii-type coupled equations for the complex parameter $\psi(x)$ and the real parameter $b(x)$
\begin{subequations}
\begin{align}
   \mu \psi (x) &= \left[ - \frac{\hbar^2}{2m} \frac{d^2}{dx^2} + V_1 (x) +  \frac{U_0 \,\mathcal{N}^2}{2\pi b^2(x) |\psi(x)|^2}
   + \frac{1}{4} b^2 (x) m \omega_r^2 + \frac{3}{4} \frac{\hbar^2}{m b^2 (x)} \right] \psi(x), \\
   b^4 (x) &= \frac{1}{m \omega_r^2} \left[ \frac{\hbar^2}{m} + \frac{1}{\pi} U_0 \,\mathcal{N}^2 |\psi(x)|^2 \right].
\end{align}
\end{subequations}
Here, $V_1(x)$ is the bare lattice potential, $U_0 = 4\pi \hbar^2 \tilde{a}/ m$, and $\omega_r$ is the radial trapping frequency. We discretize the $x$-axis into $2000$ points and iterate these equations to obtain the self-consistent solution for $\psi(x)$ and $b(x)$. 

\section{Analytic explanation for high-order phasonic multiphoton resonances}

Here we provide a qualitative explanation for the appearance of multiphoton resonances of relatively high order in the phasonic driving case. We begin by writing down a basic Hamiltonian that captures the essential features of the experiment and  describes the motion of a single particle in the potential produced by superimposing (i) a stationary  primary lattice of amplitude $V_P$ and (ii) a moving secondary lattice of amplitude $V_S$, whose position is time-dependent in a harmonic fashion with a certain amplitude $A$. The Hamiltonian is
\begin{equation}
\label{HAMp}
  H = - \frac{\hbar^2}{2m} \frac{d^2}{dx^2} + \frac{V_P}{2} \cos (2 k_P x )
  + \frac{V_S}{2} \cos \left\{2 k_S [x - A \sin (\omega t)]\right\}.
\end{equation}
We observe that the last time-dependent term is readily analyzed using the Jacobi-Anger expansion with the result
\begin{equation}
\begin{split}
  \!\!\!\cos \left\{2 k_S [x - A \sin (\omega t)]\right\} 
  %&= \cos (2 k_S x) \cos [2 k_S A \sin (\omega t)] + \sin (2 k_S x) \sin [2 k_S A \sin (\omega t)] \\
  &= \cos (2 k_S x) J_0 (2 k_S A) \\
  &+ 2 \cos (2 k_S x) \sum_{n=1}^{\infty} J_{2n} (2 k_S A) \cos (2n\omega t) \\
  &+ 2 \sin (2 k_S x) \sum_{n=1}^{\infty} J_{2n-1} (2 k_S A) \sin [(2n-1)\omega t].
\end{split}
\end{equation}
This demonstrates that the driven Hamiltonian acquires terms featuring higher temporal harmonics,
i.e.\ terms oscillating as $\sin [(2n-1)\omega t]$ or $\cos (2n\omega t)$. We note, that the 
relevance of these higher harmonics is governed by prefactors proportional to the Bessel functions
of order equal to the order of the harmonic in question and argument $2 k_S A$. Since the Bessel functions quickly decay as the argument decreases significantly below the order, it is clear that the higher harmonics can only contribute as long as $2 k_S A \gtrsim n$ or equivalently
\begin{equation}
\label{eq:condition}
   \frac{2 k_S A}{n} \gtrsim 1.
\end{equation}
Under the experimental conditions, $k_S = 6.867 \times 10^{6} \,\mathrm{m}^{-1}$, the driving amplitude 
was scaled according to 
\begin{equation}
\label{eq:expt1}
  A  =  \frac{C \, \alpha_\mathrm{phason}}{f_\mathrm{phason}},
\end{equation}
%$C=1000 \text{ nm}\cdot\text{kHz}$
with $C = 1000 \, \mathrm{nm} \cdot \mathrm{kHz}$, 
and the order of the multiphoton resonance is given by
\begin{equation}
\label{eq:expt2}
  n = \frac{f_0}{f},
\end{equation}
with $f$ denoting the drive frequency and $f_0 \approx 15 \,\mathrm{kHz}$ being the 
fundamental resonance frequency. Plugging Eqs.~(\ref{eq:expt1})--(\ref{eq:expt2}) into 
Eq.~(\ref{eq:condition}), we see that the left-hand side of the condition is independent of frequency 
and evaluates to $0.92 \, \alpha_\mathrm{phason}$, thus allowing the resonances to be observed in a broad range of frequencies
as soon as the factor $\alpha_\mathrm{phason}$ exceeds a threshold value close to unity.

\end{document}